\documentclass[conference]{IEEEtran}
\IEEEoverridecommandlockouts

\usepackage{bm}
\usepackage{graphicx}
\usepackage{balance}
\usepackage{diagbox}
\usepackage{amsmath,amssymb,amsfonts}
\usepackage{algorithmic}
\usepackage{graphicx}
\usepackage{textcomp}
\usepackage{url}
\usepackage[T1]{fontenc}

\usepackage{caption}
\usepackage{subcaption}
\usepackage{stfloats}
\usepackage{booktabs,caption}
\usepackage{hyperref}
\usepackage[ruled,linesnumbered]{algorithm2e}
\usepackage[flushleft]{threeparttable}
\usepackage{multirow}

\usepackage{array}
\newcolumntype{P}[1]{>{\centering\arraybackslash}p{#1}}
\usepackage{cite}
\usepackage{amsmath,amssymb,amsfonts}
\usepackage{algorithmic}
\usepackage{graphicx}
\usepackage{textcomp}
\usepackage{xcolor}
\def\BibTeX{{\rm B\kern-.05em{\sc i\kern-.025em b}\kern-.08em
    T\kern-.1667em\lower.7ex\hbox{E}\kern-.125emX}}
\begin{document}

\title{Towards Integrated Energy-Communication-Transportation Hub: A Base-Station-Centric Design in 5G and Beyond\\
}

\author{\IEEEauthorblockN{Linfeng Shen\IEEEauthorrefmark{2}, Guanzhen Wu\IEEEauthorrefmark{2}, Cong Zhang\IEEEauthorrefmark{3}\IEEEauthorrefmark{4}\IEEEauthorrefmark{1}\thanks{$^*$Corresponding authors.}, Xiaoyi Fan\IEEEauthorrefmark{3}\IEEEauthorrefmark{5}, Jiangchuan Liu\IEEEauthorrefmark{2}\IEEEauthorrefmark{3}\IEEEauthorrefmark{1}}
\IEEEauthorblockA{\IEEEauthorrefmark{2}School of Computing Science, Simon Fraser University}

\IEEEauthorblockA{\IEEEauthorrefmark{3}Jiangxing Intelligence Inc.}

\IEEEauthorblockA{\IEEEauthorrefmark{4}The University of Hong Kong}
\IEEEauthorblockA{\IEEEauthorrefmark{5}The Hong Kong University of Science and Technology}}

\maketitle
\begin{abstract}
The rise of 5G communication has transformed the telecom industry for critical applications. With the widespread deployment of 5G base stations comes a significant concern about energy consumption. Key industrial players have recently shown strong interest in incorporating energy storage systems to store excess energy during off-peak hours, reducing costs and participating in demand response. The fast development of batteries opens up new possibilities, such as the transportation area. An effective method is needed to maximize base station battery utilization and reduce operating costs. In this trend towards next-generation smart and integrated energy-communication-transportation (ECT) infrastructure, base stations are believed to play a key role as service hubs. By exploring the overlap between base station distribution and electric vehicle charging infrastructure, we demonstrate the feasibility of efficiently charging EVs using base station batteries and renewable power plants at the Hub. Our model considers various factors, including base station traffic conditions, weather, and EV charging behavior. This paper introduces an incentive mechanism for setting charging prices and employs a deep reinforcement learning-based method for battery scheduling. Experimental results demonstrate the effectiveness of our proposed ECT-Hub in optimizing surplus energy utilization and reducing operating costs, particularly through revenue-generating EV charging.
\end{abstract}


\section{Introduction}

In recent years, the rapid development of 5G communication has sparked a transformation in the telecommunications industry. Due to its unmatched speed and connectivity, 5G is widely recognized as the preferred technology for critical applications like autonomous driving \cite{devi20235g}, augmented reality, and wearable devices \cite{chekired20195g}. The adoption of 5G in these applications is driven by its capability to support high-speed, low-latency communication, enabling real-time interactions between various devices and systems. To offer a greater coverage area and service quality, the deployment of 5G networks involves a large number of base stations (BSs). According to a statistical report, as of October 2022, more than 3 million BSs have been deployed globally \cite{web}. The rapid expansion of 5G infrastructure also poses challenges, particularly in terms of energy consumption and operating costs. 5G BSs exhibit substantial energy demands, and their power consumption escalates correspondingly with the increase in data traffic \cite{piovesan2022power}. In the UK, for example, power consumption from 5G networks is expected to reach 2.1\% of total UK-wide electricity generation in 2030 under medium load demand \cite{cheng20225g}. Moreover, these BSs need to be equipped with backup batteries to provide uninterrupted service during a blackout for some critical applications such as autonomous driving. The management costs and the self-degradation process of batteries \cite{8370662} impose additional burdens on operators. Effectively utilizing these backup energies remains a challenge for BS operators.

To take advantage of the energy storage capacity of batteries, researchers have explored some solutions to optimize energy usage and reduce operational costs. One effective approach is the integration of backup batteries as Battery Energy Storage Systems (BESS) into the 5G infrastructure \cite{tang2021reusing}. BESS can store excess energy during off-peak hours and release it during peak demand periods, balancing the fluctuations in energy supply and demand. Operators can save costs by utilizing BESS and taking advantage of off-peak electricity pricing. The energy consumption of a single base station is relatively small compared to the typical storage battery capacity used in grid-scale applications (hundreds of KWh). Additionally, batteries undergo self-degradation even when not in use, potentially leading to waste. There is still huge potential for BESS. Feeding power back from BESS to the grid or microgrid, although a common idea, is not feasible due to the fluctuations in grid integration that disrupt the normal operation of the grid, resulting in risks and expenses that far outweigh the benefits of grid integration \cite{delgado2015smart}.

\begin{figure}[t]
  \centering
  \subcaptionbox{Main roads\label{fig3:a}}
  {\includegraphics[width=0.45\linewidth]{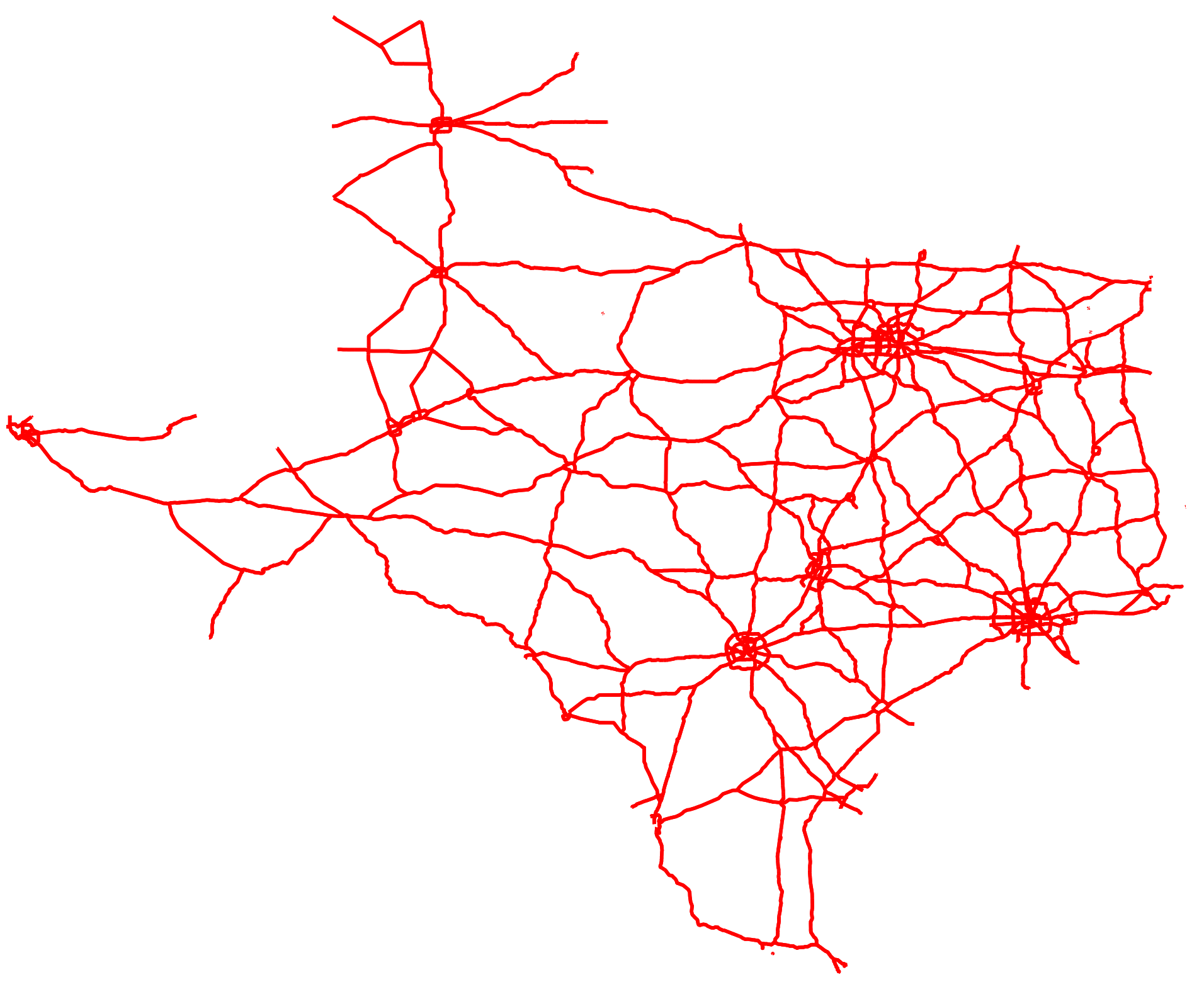}}\hspace{0.2em}%
  \subcaptionbox{Base stations\label{fig3:a}}
  {\includegraphics[width=0.45\linewidth]{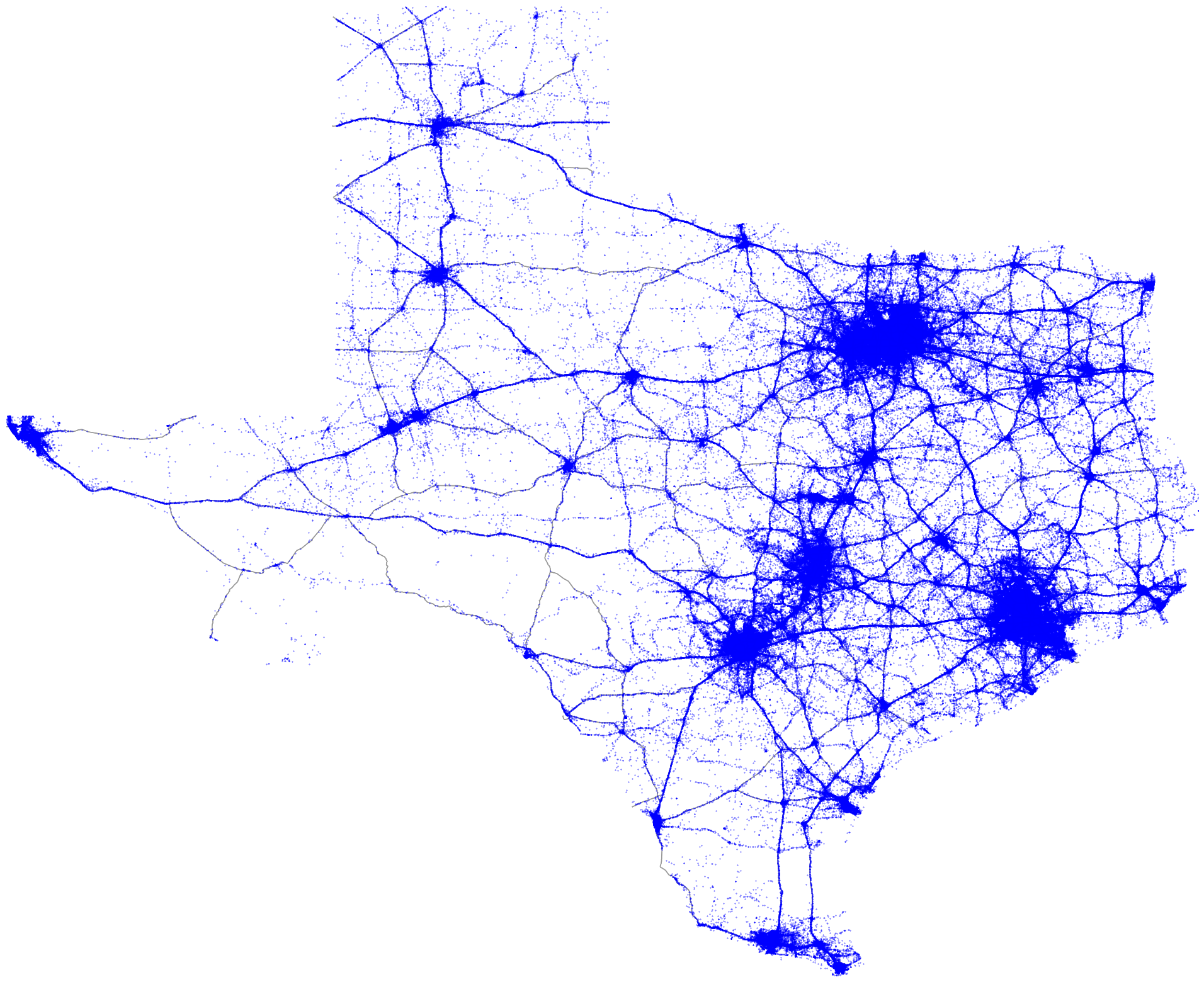}} 
  \caption{The distribution of main roads and base stations in Texas, USA. The roads and base stations are represented by the red lines and the blue points\cite{road,opencellid}.}
  \label{road}
\end{figure}

On the other hand, electric vehicles (EVs) are becoming popular transportation choices as battery technology and charging infrastructure improves, and an EV can be seen as a mobile load during the charging process. Considering the new concept of Internet-of-Vehicles (IoV) \cite{ni2022lagrange} and services closely combine BSs and EVs in 5G and future 6G, such as in-vehicle communication and autonomous driving, the driving traces of EVs should highly overlap with the distribution of BSs. Fig.~\ref{road} illustrates the distribution of main roads and BSs in Texas, showcasing their high degree of coincidence. This distribution suggests a likely increase in the density of BS deployments alongside roads in the 5G and forthcoming 6G eras. It is reasonable to expect that future deployments of more BSs alongside roads will make it convenient to charge EVs with excess energy from BSs. Due to the limited space and infrastructure of EVs (the size of the DC/AC converter is limited), charging from BESS in BSs is more efficient than the charging stations since they both use DC power \cite{safayatullah2022comprehensive}, which benefits both the customers and operators of BSs. Prominent industry leaders are actively exploring the integration of energy, communication, and transportation services. Elon Musk has publicly discussed his vision of integrating Tesla's transportation services with SolarCity's energy solutions and Starlink's communication capabilities\footnote{https://www.tesla.com/blog/master-plan-part-deux}\footnote{https://www.linkedin.com/pulse/elon-musks-innovative-approach-vertical- integration-closed/}. Similarly, China Tower, a leading telecommunications provider in China, has been deploying 5G base stations on the existing transmission tower for years\footnote{https://www.china-tower.com/Index/show/catid/19/id/1086.html}, and also exploring energy services leveraging the power capabilities of base stations, such as battery swapping for electric bikes\footnote{https://www.china-tower.com/Index/lists/catid/77.html}. This trend toward integration positions 5G base stations as key hubs in the energy, communication, and transportation integration design. In this work, we investigate the feasibilities and challenges of energy-communication-transportation hub (ECT-Hub) design from a base-station-centric view and propose methods to tackle the challenges while maximizing the profit of ECT-Hub operators.

The introduction of charging functions and BESS to the BSs adds complexity to the system and presents new challenges. At first, the primary task is to ensure the BSs maintain basic communication functions. Then, a number of the BSs have already been equipped with renewable power sources, e.g., wind or solar \cite{matracia2021exploiting}, which will be common in the future network infrastructure. To optimize system costs, scheduling the operation of BESS to handle fluctuating renewable power generation and varying power prices presents an additional challenge. Another challenge is determining optimal charging prices at each ECT-Hub to attract more EVs and maximize profits. Effective personalized incentives are potential solutions for the charging price decision. We employ a causal inference-based method \cite{li2023counterfactual} to determine optimal times to offer charging price discounts, aiming to maximize profit. Meanwhile, we use a deep reinforcement learning (DRL) method to manage the operation of BESS in the system, considering the operation cost of each part. In collaboration with a major cellular service provider in China, we gather backup battery data from BSs and charging records from electric vehicles. The experiments demonstrate the system's effectiveness, with our methods outperforming the baseline comparisons. The main contributions of this work are summarized as follows:

\begin{itemize}
\item We systematically investigate an integrated energy-communication-transportation hub design from a base-station-centric view. Without sacrificing the communication service quality, we demonstrate the feasibility and potential of such a design and identify the key challenges in supporting EV charging.
\item We introduce a novel causal inference method to regulate the
charging price in realtime, together with smart scheduling of battery allocation and operation in the system. Under stringent communication demands, our solution maximizes the profit of operators and well accommodates renewable power sources. 
\item Working with key industrial players from both the communication and energy sectors, we have evaluated our solution with representative public datasets
and up-to-date proprietary datasets. The results demonstrate the effectiveness of the integrated design and the superior performance of our methods, in particular, \textbf{ECT-Price} and \textbf{ECT-DRL} as compared to the baseline.
\end{itemize}

\section{Feasibility and Challenges}

In this section, to verify the feasibility of the ECT-Hub design from the base-station-centric view, we first make some measurements based on real-world datasets. The measurement results not only demonstrate the feasibility of the design and potential benefits but also expose some challenges during the operation. To tackle these challenges, we propose specific methods in this work and try to maximize the profit for the operators of ECT-Hubs.
\subsection{Technology Feasibility}
As shown in Fig.~\ref{road}, we have demonstrated a significant correlation between the distribution of base stations and road infrastructure. In some contexts, the location of base stations is ideal for charging EVs, such as the underground parking lot where the base stations are deployed on top of the building or the base station towers alongside the rural roads. Although most base stations are deployed on the top of buildings (especially in urban areas) for better coverage, the backup batteries are usually located in the specific battery room on the ground for better management. These backup batteries are connected to the base station with a DC power system and automatic transfer switch to provide power in emergencies. We propose transforming base stations into energy-communication-transportation integrated hubs by adding electric vehicle supply equipment (EVSE), which can utilize excess energy from base station batteries or renewable power generators. We aim to ensure uninterrupted communication while optimizing battery usage and incorporating renewable energy generation. This approach involves designing a comprehensive battery management and charging pricing system to maximize benefits for base station operators. 

Battery pack capacities have significantly increased since 2011 (200kWh-600kWh), far surpassing the needs of a single 5G base station (2-4kW) \cite{6695255} nowadays. Considering the better efficiency and less environmental damage, there is a trend to replace current lead-acid batteries in base stations with lithium-ion batteries, which are also the major ones used in EVs. Using retired EV lithium-ion batteries in base stations is also proposed and already adopted by some operators \cite{yang2020environmental}. To fully use the capacity storage of batteries and save base station operators' expenses,  storage batteries in the base station have been commercialized as battery swapping stations to provide power exchange for electric bikes \cite{twowheeler}. However, several issues persist in EV battery swapping, such as AC to DC conversion and standardizing batteries across different brands. The prevailing approach remains building charging stations for EV recharging. Previous work \cite{yan2017catcharger} also proposes to use wireless power transfer (WPT) to charge EVs. The cost of WPT is still very high, while the efficiency is quite low compared to wired power transfer, and some research \cite{lin2022you} has pointed out the security issues of the current WPT technology. To this end, we only focus on directly charging EVs with EVSE in the base-station-centric ECT-Hub design. Since the ECT-Hub can provide both AC power from the connected grid and DC power from equipped batteries, the EVSE installed in the ECT-Hub can provide both AC dedicated and DC dedicated interface defined in the IEC 62196 standard \cite{standard} without rectifiers, which can further improve the energy efficiency for the whole system.

\begin{figure} 
  \centering 
  \begin{minipage}{0.49\linewidth} 
    \centering
    {\includegraphics[width=\linewidth]{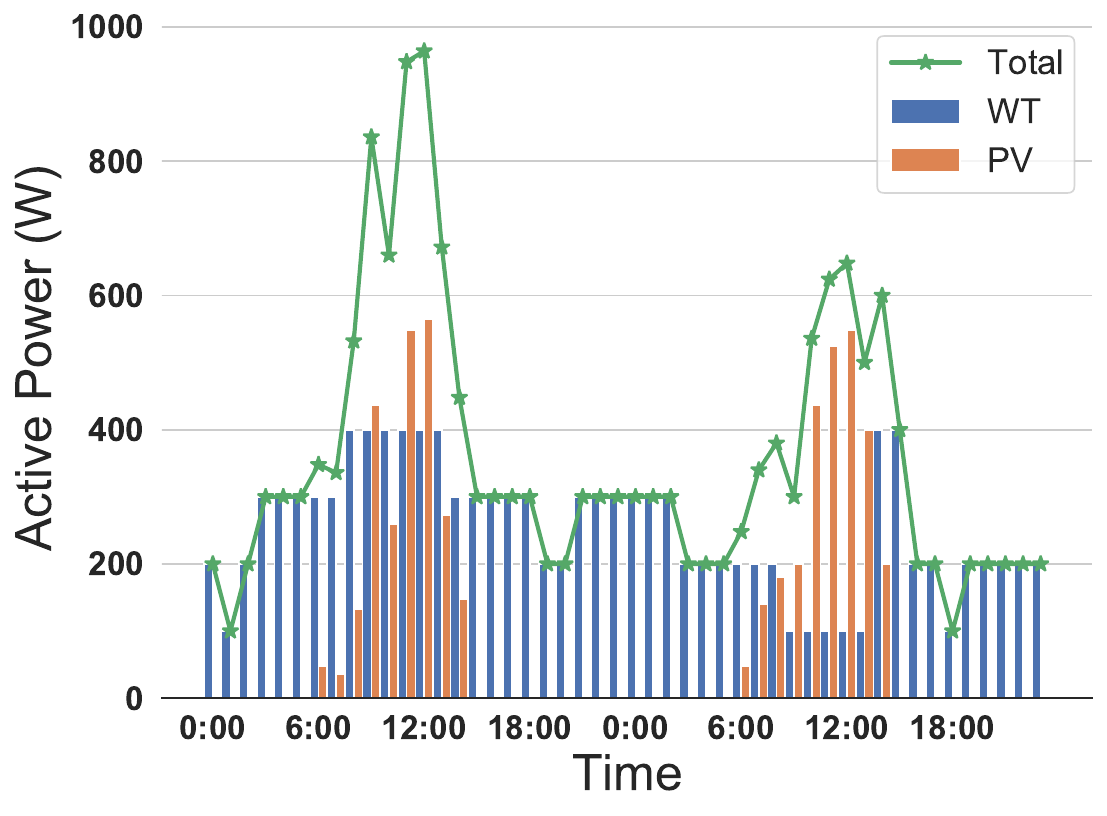}} 
    \captionof{figure}{Active power of renewable power generation.} 
    \label{fig:res_power}  
  \end{minipage} 
  \begin{minipage}{0.49\linewidth} 
    \centering 
{\includegraphics[width=\linewidth]{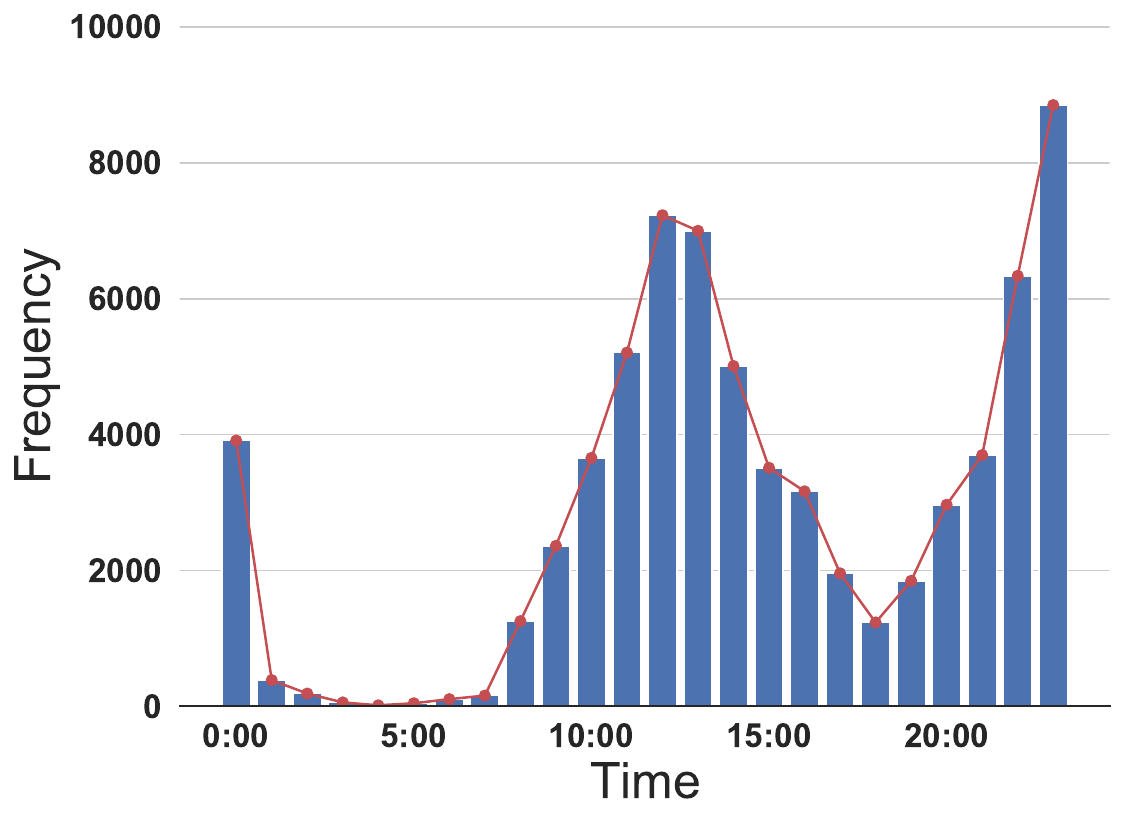}} 
    \captionof{figure}{Charging frequencies of electric vehicles.} 
    \label{fig:charge_freq} 
  \end{minipage} 
  \end{figure}

\subsection{Economic Feasibility}

The introduction highlights that electricity cost is a significant expense for 5G network operators. A typical 5G base station can consume three to four times more energy than a 4G base station \cite{tang2021reusing}. The energy consumption of 5G base stations is positively correlated with the load rate. A 5G base station primarily consists of the Baseband Unit (BBU) and the Active Antenna Unit (AAU). We note that the energy consumption of the BBU remains relatively stable, while the AAU's energy consumption increases with the load rate. Thus, the network traffic is a good indicator for predicting electricity costs for 5G base stations.

Introducing renewable energy generation (such as wind and solar power) and energy storage solutions (batteries) in base station construction is a promising approach to reduce electricity expenses for 5G operators. Some base stations, especially those in remote areas, are equipped with wind and solar power systems \cite{peng2022optimal}. As shown in Fig.~\ref{fig:res_power}, for two days of wind and photovoltaic power generation, renewable power generation has great volatility and is hard to predict in advance. The current base station network cannot fully utilize this power, leading to energy wastage. On the other hand, the charging requirement of EVs also has different characteristics during one day. For example, Fig.~\ref{fig:charge_freq} depicts the real-world data from 12 charging stations over 3 years with 70,000 records, showing significant usage variation at different times. With appropriate methods to balance the charging requirement of EVs and the power generation, the operators of ECT-Hubs can fully utilize the available energy and make more profit.

\begin{figure} 
  \centering 
  \begin{minipage}{0.49\linewidth} 
    \centering    
    {\includegraphics[width=\linewidth]{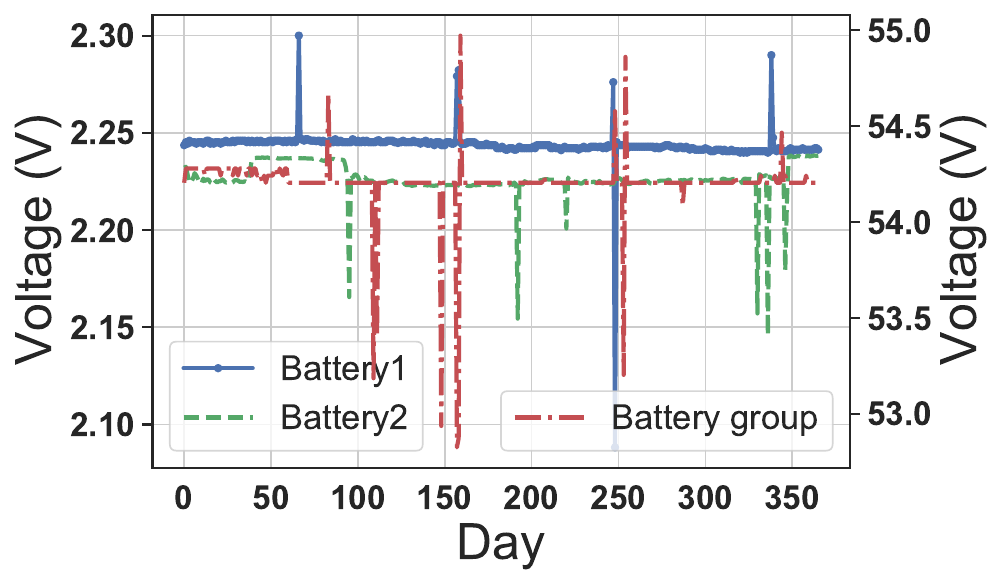}} 
    \captionof{figure}{Voltage of two batteries and a battery group.} 
    \label{fig:voronoi} 
  \end{minipage} 
  \begin{minipage}{0.49\linewidth} 
    \centering 
    {\includegraphics[width=\linewidth]{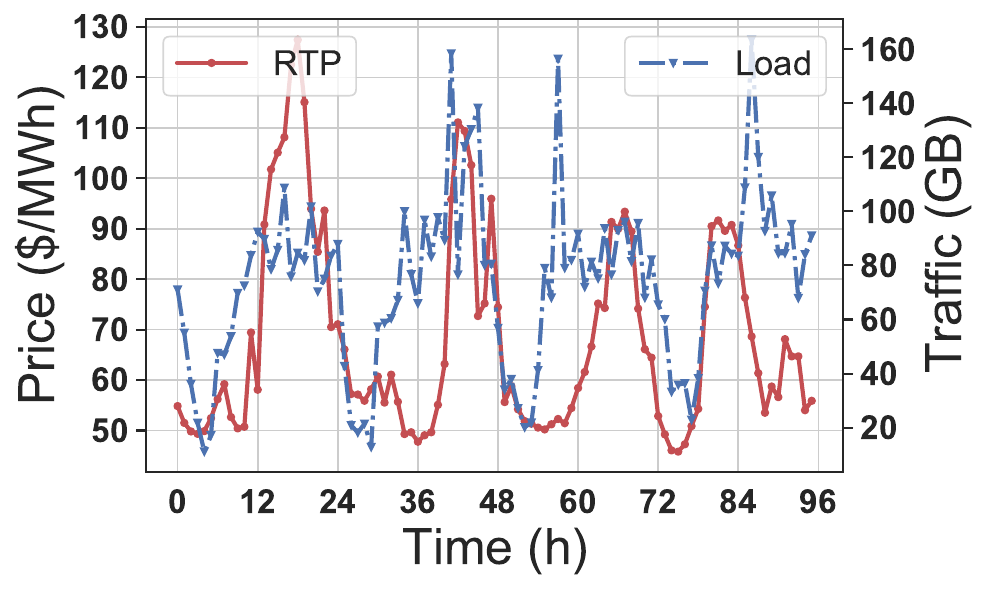}} 
    \captionof{figure}{Real-time pricing and network traffic.} 
    \label{fig:illustrate_cases} 
  \end{minipage} 
  \end{figure}

Battery degradation is another cost for the operators. Backup batteries will eventually degrade even without usage. Battery voltage is an important indicator for evaluating the state of the battery. Fig.~\ref{fig:voronoi} demonstrates the capacity degradation of two individual batteries and a battery group over time~\cite{8370662}. It can be seen that the voltage of the battery gradually decreases with time, reflecting the slow degradation process of the battery. If we can utilize these backup energies to charge EVs at ECT-Hub while guaranteeing the basic communication functions, the operation cost caused by battery degradation can be neutralized somewhat.

\subsection{Challenges of ECT-Hub}

Although we have introduced many benefits of the ECT-Hub and its feasibility, some challenges still need to be considered during the design of the ECT-Hub. At first, the basic communication functions of the base stations must not be interrupted while minimizing operating costs. The network traffic in a region, however, can fluctuate and be hard to predict. In Fig.~\ref{fig:illustrate_cases}, we presented the results of our measurement study, where we analyzed four days of network traffic data along with real-time electricity price data from a public dataset \cite{Modeling15-Chen}. We observe that the load rate of base stations is positively correlated with the electricity price. Notably, both load factors and electricity prices peak during the night, leading to a significant increase in electricity expenses for these base station operators. Energy storage systems and renewable power generations can be utilized to balance energy. However, it is crucial to acknowledge that incorporating batteries and renewable energy sources comes with significant challenges, namely the degradation of batteries and the uncertainty of new energy generation. Previous work~\cite{tang2021reusing} proposed using base station backup batteries as BESS to optimize energy expenditure through battery scheduling. However, this work did not account for integrating renewable energy sources. Thus, we need to design an effective method to schedule the batteries in the ECT-Hub given all these challenges.

On the other hand, the operators want to make more profit by charging EVs with extra energy, either stored in backup batteries or generated by renewable power plants. The willingness of EV owners to charge is highly influenced by the price of electricity, which is set by the operators. This prompts the crucial question of how to optimize charging service pricing to maximize revenue. Dynamic pricing for charging services is essential due to the fluctuating demands for EV charging throughout the day, as shown in Fig.~\ref{fig:charge_freq}. An efficient pricing strategy, synchronized with peak and off-peak periods, is needed to optimize revenue for operators and encourage EV owners to charge during less congested hours, balancing loads and enhancing network stability.

\section{System Model}
\begin{table}[t!]
\vspace{0.1cm}
\caption{Notations.}
\centering
 \begin{tabular}{p{40pt}|p{160pt}}
 \hline
\textbf{Notation}& \textbf{Description}\\
 \hline
 ${t_1,t_2,...,t_{\mathbb{T}}}$ & time slots   \\
 $T_r$ & estimated recovery time of the power grid \\
 $\alpha_t$ & load rate of base station at time $t$\\
 $\eta_{ch(dch)}$ & charging (discharging) efficiency of BP\\
 $c_{BP}$ & operation cost of BP for each time slot\\
 $R_{ch(dch)}$ & charging (discharging) rate of BP \\
 \hline
 $P_{BS}^{max}$ & power of base station with full load rate \\
 $P_{BS}^{min}$ & power of base station with idle state\\
 $P_{BS}(t)$ & power consumption of base station at time $t$\\
 $P_{CS}(t)$ & power consumption of charging station at time $t$\\
 $P_{BP}(t)$ & power consumed
or provided by the BP at time $t$\\
$P_{WT}(t)$ & power generated by the wind turbine at time $t$\\
$P_{PV}(t)$ & power generated by the photovoltaic at time $t$\\
$P_{grid}(t)$ & power required from the power grid at time $t$\\
 \hline
 $S_{CS}(t)$ & state of the charging station at time $t$   \\
 $S_{BP}(t)$ & state of the BP at time $t$   \\
 \hline
 $RTP(t)$ & real-time price of electricity at time $t$ \\ 
 $SRTP(t)$ & selling price of electricity through CSs at time $t$\\
 \hline
 $SoC(t)$ & state of charge (SoC) of BP at time $t$ \\
 $SoC_{min}$ & lower bound SoC of BP\\
 $SoC_{max}$ & upper bound SoC of BP\\
 \hline
 $C_{BP}(t)$ & cost of BP at time $t$ \\
 $C_{grid}(t)$ & cost of power grid at time $t$ \\
 \hline
 \end{tabular}
\end{table}
In this section, we first present the system model of the base-station-centric ECT-Hub and introduce the relationship of each part in the system. Then, we provide the problem formulation, considering the objective of ECT-Hub operators. The notations used in this paper are shown in Table I.

\subsection{System Overview}

\begin{figure}[tb!]
\centerline{\includegraphics[width=0.8\linewidth]{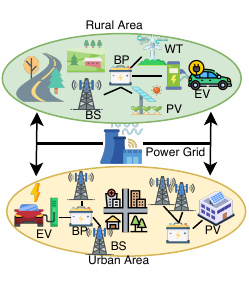}}
\vspace{-0.5cm}
\caption{System overview.}
\label{system}
\end{figure}

Fig.~\ref{system} shows the overview of the proposed system. We consider reconstructing base stations into ECT-Hubs, which are equipped with renewable power generation plants and charging stations for electric vehicles, in addition to basic communication functions in the future 5G network and beyond. Nowadays, most base stations are already equipped with battery groups as backup power supplies. The battery groups of one base station or several nearby base stations can serve as the battery points (BPs) for the ECT-Hubs and compose a BESS, which can help to save operating expense (OPEX) with scheduling \cite{tang2021reusing}. The distribution of base stations is very dense and the charging demands are also huge in the urban area. Renewable power generation plants, however, are hard to deploy due to the limited space in urban areas, where the photovoltaic (PV) power plants installed on the top of buildings are promising solutions. On the other hand, renewable power resources are much more abundant in rural areas, where both large-scale PV and wind turbine (WT) power plants are feasible. The charging demands are also different in rural areas. All these factors will influence the operating cost of the ECT-Hubs. How to schedule the BESS and decide the selling electricity price of charging stations to maximize the profit remains a big challenge for the ECT-Hubs providers. The rest of this section introduces the details of each part.

\subsection{Design Details}
As shown in Fig.~\ref{system}, the power supply of BSs can be the power grid directly or the equipped BP. Some ECT-Hubs are also equipped with PV or WT power plants, which can provide extra but irregular power. Previous works \cite{piovesan2022power} \cite{tang2021reusing} have shown that the power consumption of 5G base stations has an approximately linear relationship with the load rate. Let $P_{BS}^{max}$ and $P_{BS}^{min}$ denote the power of BSs in idle state and with full load rate, respectively. The power of one BS at time $t$ with load rate percentage $\alpha_t$ can be represented as:

\begin{equation}
P_{BS}(t) = P_{BS}^{min}+ \alpha_t(P_{BS}^{max}-P_{BS}^{min}),
\label{power_BS}
\end{equation}
where ${t_1,t_2,...,t_\mathbb{T}}$ represents the time slots in a time period $\mathbb{T}$, and $P_{BS}(t)$ is the power consumption of BS at time $t$. 

Another power consumer in the ECT-Hub is the charging stations (CSs) for EVs. The power supply of CSs is either the
power grid directly or the equipped BP. Let $S_{CS}(t)$ denote the state of the CS at time $t$. $S_{CS}(t) = 1$ if there are EVs charging at time $t$, otherwise $S_{CS}(t) = 0$. The power consumption of the CS at time $t$ can be represented as:

\begin{equation}
P_{CS}(t)=S_{CS}(t)\cdot R_{CS},
\label{power_CS}
\end{equation}
where $R_{CS}$ is the charging rate of EVs.

Previous works \cite{tang2021reusing} \cite{ccimen2020microgrid} have shown that BESS can not only serve as backup power during a blackout but also provide opportunities for demand reshaping to save cost in both industrial and residential scenarios. In this case, the $BP$ can serve as the power provider when discharging to other parts of the ECT-Hub and the power consumer when charging from the power grid as well. Let $P_{BP}(t)$ denote the power consumed or provided by the BP at time $t$, which can be calculated as:

\begin{equation}
P_{BP}(t) = S_{BP}(t)\cdot\eta_{ch(dch)}\cdot R_{ch(dch)},
\label{power_BP}
\end{equation}
where $\eta_{ch(dch)}$ is the charging (discharging) efficiency of BP, $R_{ch(dch)}$ is the charging (discharging) rate of BP, and $S_{BP}(t)$ represents the state of BP. $S_{BP}(t)=1$ if BP is charging, $S_{BP}(t)=-1$ if BP is discharging, and $S_{BP}(t)=0$ if BP is not used. 

The level of charge of the batteries relative to their capacity is usually measured by the State of Charge (SoC). The SoC of the $BP$ at time $t+1$ can be calculated as

\begin{equation}
SoC(t+1)=SoC(t)+P_{BP}(t).
\label{soc}
\end{equation}

To relieve the degradation process of the batteries \cite{8370662}, the SoC of BP should also be limited in a certain lower and upper bound to have a longer lifetime, as shown in Eq. \ref{soc_bound}. On the other hand, the lower bound of SoC should also guarantee the basic operation of base stations during blackout before the recovery of the power grid, as shown in Eq. \ref{soc_lowerbound}:

\begin{equation}
SoC_{min}\leq SoC(t)\leq SoC_{max},
\label{soc_bound}
\end{equation}
\begin{equation}
\sum_{t}^{t+T_r}P_{BS}(t)\leq SoC_{min},
\label{soc_lowerbound}
\end{equation}
where $SoC_{min}$ and $SoC_{max}$ represent the lower bound and upper bound SoC of BP. $T_r$ is the estimated recovery time of the power grid.

The power supply of ECT-Hub can also be from renewable power generation plants or the power grid directly. If the ECT-Hub is equipped with WT or PV power plants, the power generated at time $t$ is represented as $P_{WT}(t)$ and $P_{PV}(t)$, respectively. The values of $P_{WT}(t)$ and $P_{PV}(t)$ are based on the weather conditions (wind speed and solar radiation) at time $t$. If the power provided by the BP and renewable power generation plants is not enough for the ECT-Hub, the power grid can provide the extra power directly. The power required from the power grid at time $t$ can be calculated as:

\begin{equation}
\begin{aligned}
P_{grid}(t)&=max\{0,P_{BS}(t)+P_{CS}(t)+P_{BP}(t)\\&-P_{WT}(t)-P_{PV}(t) \}.
\end{aligned}
\label{pgrid}
\end{equation}

\subsection{Problem Formulation}
The objective of the ECT-Hubs operator is to maximize profit by selling extra power to EVs through integrated charging stations. However, the maintenance of BPs and the power required from the power grid introduce additional costs. To measure the extra cost caused by the degradation of batteries, we use $C_{BP}(t)$ to represent the operation cost of BP at time $t$, which is shown as Eq. \ref{soc_cost}, where $c_{BP}$ is the operation cost of BP for each time slot.

\begin{equation}
C_{BP}(t)=|S_{BP}(t)|\cdot c_{BP}.
\label{soc_cost}
\end{equation}

The power required from the power grid incurs another cost considered in our system. Let $RTP(t)$ represent the real-time price of electricity at time $t$. This part of costs can be represented as:

\begin{equation}
C_{grid}(t)=P_{grid}(t)\cdot RTP(t).
\label{grid_cost}
\end{equation}

The total operation cost $OC$ can be calculated as the summation of money paid for power from the power grid and the cost of BP over the whole time period $\mathbb{T}$:

\begin{equation}
OC=\sum_{t\in \mathbb{T}}\left[ C_{grid}(t)+ C_{BP}(t)\right].
\label{oc}
\end{equation}

On the other hand, the ECT-Hubs operator can determine the selling price for the charging EVs to maximize the profit. This part of the reward can be calculated as:

\begin{equation}
CR=\sum_{t\in \mathbb{T}} \left[ P_{CS}(t)\cdot SRTP(t)\right],
\label{CR}
\end{equation}
where $SRTP(t)$ represents the selling price of electricity through CSs at time $t$.

Based on the above modeling of each part in the system, the objective of the proposed ECT-Hubs is to maximize the overall profit, which can be calculated as:
\begin{equation}
\begin{aligned}
\textbf{Max:}~\Psi&= CR - OC\\
&=\sum_{t\in \mathbb{T}} \{ P_{CS}(t)\cdot SRTP(t) - P_{grid}(t)\cdot RTP(t) \\
&- |S_{BP}(t)|\cdot c_{BP} \},
\end{aligned}
\label{obj}
\end{equation}
where the scheduling of BP $S_{BP}(t)$ and the charging price $SRTP(t)$ are two variables that are determined by the ECT-Hub itself, while the specific environment gives all the others and cannot be handled by the operator.
\section{Proposed Methods}

\begin{figure}[tb!]
\centerline{\includegraphics[width=\linewidth]{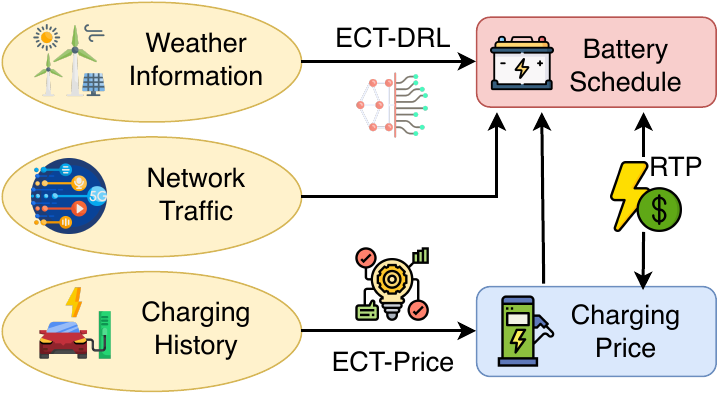}}
\caption{Framework of the proposed methods.}
\label{framework}
\end{figure}

According to the modeling of each part of ECT-Hubs and the problem formation in Section III, we propose methods to dynamically determine the operation of BPs and the charging price at each time slot. The framework of the proposed methods is shown in Fig.~\ref{framework}. The methods consist of two stages. At first, the charging price is determined based on the charging history of each charging station. To attract more EVs to charge at the ECT-Hubs, we use a causal inference-based method \textbf{ECT-Price} to determine when to give discounts on charging prices for each ECT-Hub. Next, considering the dynamic BS power demand demands in practice, deep reinforcement learning (DRL) is commonly used for the scheduling of BESS\cite{tang2021reusing}. We also propose a DRL-based method \textbf{ECT-DRL} to schedule the operation of the batteries in each ECT-Hub given the real-time electricity price, weather information, network traffic history, and the determined charging price. The rest of this section gives the details of each method.

\subsection{Charging Price Discount}

\begin{figure}[tb!]
\centerline{\includegraphics[width=0.8\linewidth]{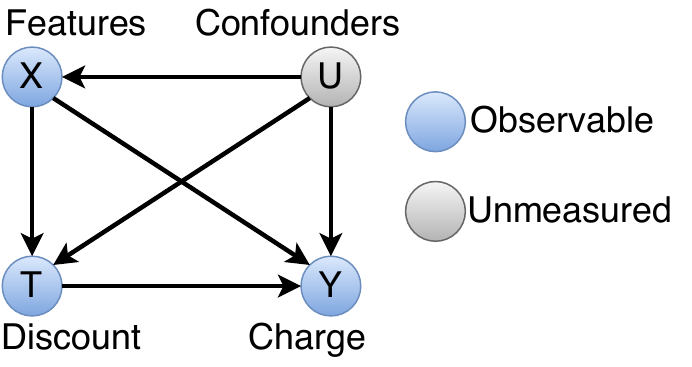}}
\caption{The causal framework of charging process.}
\label{causal}
\end{figure}

In e-commerce, effective personalized incentives can improve the user experience and increase provider revenue. Similarly, offering price discounts is expected to attract more EVs to charge at ECT-Hubs. To determine when to give price discounts, we propose a causal inference-based method \textbf{ECT-Price} based on the \textbf{CF-MTL} model \cite{li2023counterfactual}, proven effective in e-commerce. Unlike traditional machine learning methods, which only measure the correlation of data, causal inference tries to infer the causal relationship among variables. The causal framework of the charging process is shown in Fig.~\ref{causal}. The charging station and time slot features are represented as variable $X$, and another treatment variable $T$ represents the price discount given to the charging station at the specific time slot. $T=1$ means the discount is given, and $T=0$ means the discount is not given. The input features $X$ and the possible incentives $T$ determine the probability of the results $Y$, which means if there are any EVs charging at the time. $Y=1$ means the EV will charge and $Y=0$ means no charge operation. All of these variables are observable in our situation. There are also some unmeasured features $U$, such as weather and holiday information, that may have impacts on the other variables. Our objective is to give discounts $T$ to the charging station at an appropriate time slot with input features $X$, and maximize the average effect of the treatment $T$, which is calculated from the results $Y$.

Traditional causal inference methods use uplift modeling to estimate the average treatment effect, which is not effective since they only predict the impact of treatment and cannot distinguish the ``Always Buyer" \cite{li2023counterfactual}. ``Always Buyer" in our situation means that there are always EVs charging at this time, giving price discounts to those time slots will definitely decrease the overall profit of the ECT-Hub since these EVs will still charge without the discounts. To this end, we utilize the multi-task learning approach with counterfactual identification proposed in \cite{li2023counterfactual} to determine the charging price discount. 

We divide all the charging operations into three strata: \textit{Always Charge}, \textit{Incentive Charge} and \textit{No Charge}. Let $Y(T)$ denote the result when given incentive $T$ ($1$ or $0$). The probability that the input features $X$ belong to \textit{Always Charge} can be represented as $P(Y_{11}|X)=P[Y(0)=1,Y(1)=1|X]$, which means there will always be EVs charging no matter if the discounts are given. Similarly, the probability of \textit{Incentive Charge} can be represented as $P(Y_{01}|X)=P[Y(0)=0,Y(1)=1|X]$, which means there will be EVs charging only when discounts are given. The probability of \textit{No Charge} can be represented as $P(Y_{00}|X)=P[Y(0)=0,Y(1)=0|X]$, which means there will be no EVs charging no matter if the discounts are given. From the view of counterfactual identification, only \textit{No Charge} can result in the observation $(Y=0, T =1)$, and only \textit{Always Charge} can result in the observation $(Y=1, T=0)$. On the other hand, both \textit{Incentive Charge} and \textit{Always Charge} can result in the observation $(Y=1, T=1)$, and both \textit{Incentive Charge} and \textit{No Charge} can result in the observation $(Y=0, T=0)$. From this counterfactual identification process, we have:

\begin{equation}
P(Y=0|T=1,X) = P(Y_{00}|X),
\label{P1}
\end{equation}
\begin{equation}
P(Y=1|T=0,X) = P(Y_{11}|X),
\label{P2}
\end{equation}
\begin{equation}
P(Y=1|T=1,X) = P(Y_{01}|X) + P(Y_{11}|X),
\label{P3}
\end{equation}
\begin{equation}
P(Y=0|T=0,X) = P(Y_{00}|X) + P(Y_{11}|X).
\label{P4}
\end{equation}

\begin{figure}[tb!]
\centerline{\includegraphics[width=0.9\linewidth]{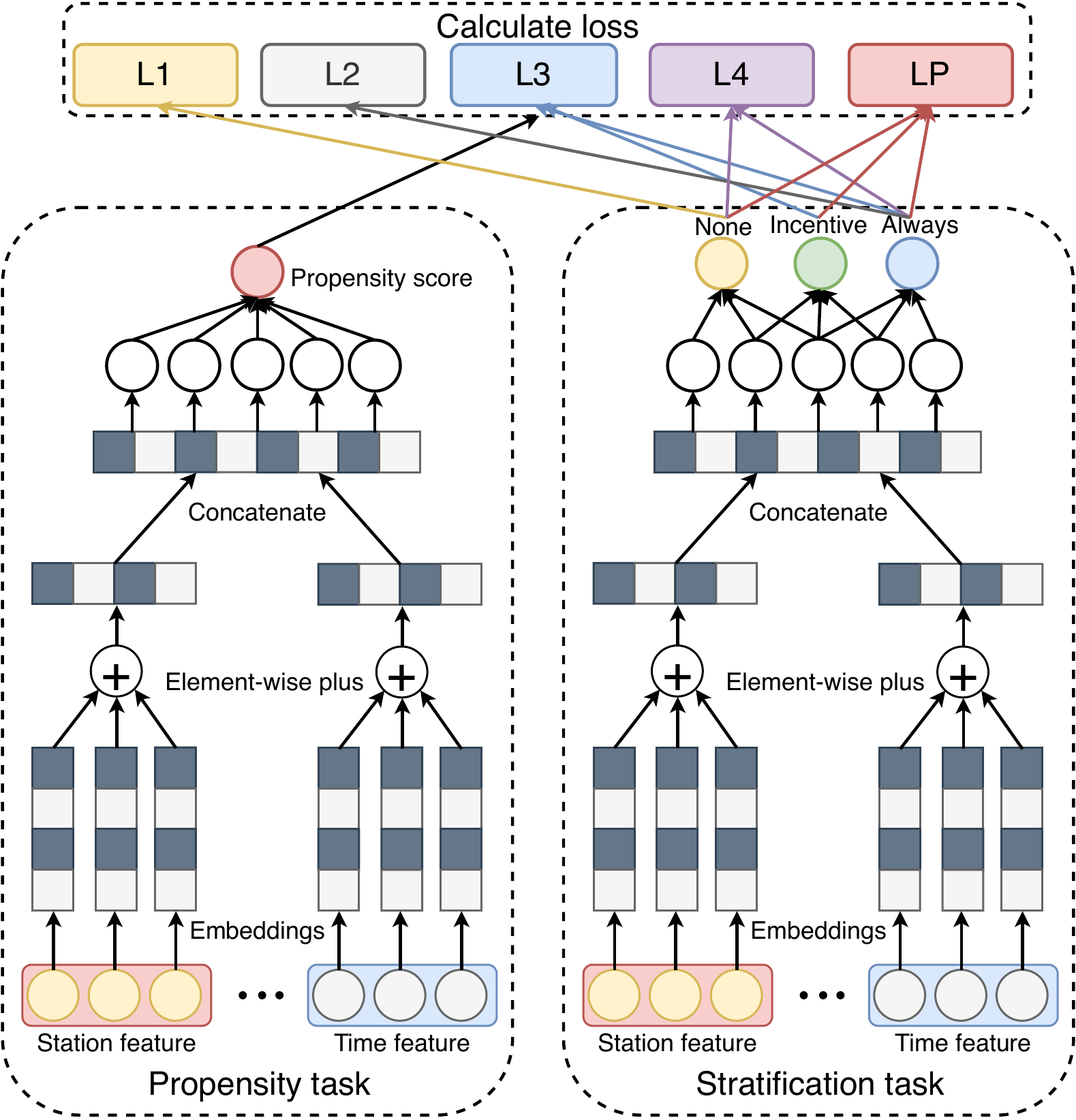}}
\caption{The architecture of ECT-Price.}
\label{CT}
\end{figure}

Based on the counterfactual identification results, we use the multi-task learning approach to predict the probabilities of each stratum. As shown in Fig.~\ref{CT}, \textbf{ECT-Price} also has a propensity task to predict the propensity score along with the stratification task. The propensity score $P(T=1|X)$ is the probability of giving incentives to input $X$. By multiplying the propensity on both sides of the Eq.~\ref{P1}, we have:

\begin{equation}
\begin{aligned}
P(Y=0|T=1)P(T=1|X) &= P(Y_{00}|X)P(T=1|X)\\
P(Y=0,T=1|X) & = P(Y_{00}|X)P(T=1|X),
\end{aligned}
\label{multi}
\end{equation}
where the left is the joint distribution of the samples with observations $(Y=0,T=1)$ and the right consists of the results of both tasks. Let $f_{00}(X)$ denote the component of the stratification task model for predicting \textit{No Charge}, and $g(X)$ denote the propensity task model. Then both models can be trained jointly by minimizing the following loss:

\begin{equation}
\mathcal{L}_1 = L(f_{00}(X)g(X), Y=0\&T=1),
\label{L1}
\end{equation}
where $L(\cdot,\cdot)$ is the average MSE loss over all pairs. $Y=0\&T=1$ equals to $1$ only when the sample has observation $(Y=0,T=1)$, and equals to $0$ otherwise. Similarly, the losses of other observations can be calculated as:

\begin{equation}
\mathcal{L}_2 = L(f_{11}(X)(1-g(X)), Y=1\&T=0),
\label{L2}
\end{equation}
\begin{equation}
\mathcal{L}_3 = L((f_{01}(X)+f_{11}(X))g(X), Y=1\&T=1),
\label{L3}
\end{equation}
\begin{equation}
\mathcal{L}_4 = L((f_{00}(X)+f_{11}(x))(1-g(X)), Y=0\&T=0).
\label{L4}
\end{equation}

In addition, the loss of the propensity task is calculated as:

\begin{equation}
\mathcal{L}_p = L(g(X), T=1).
\label{LP}
\end{equation}

At last, the total loss of \textbf{ECT-Price} is represented as follows and is minimized during the training of two tasks:

\begin{equation}
\begin{aligned}
\mathcal{L} &= \mathcal{L}_1+\mathcal{L}_2+\mathcal{L}_3+\mathcal{L}_4+\mathcal{L}_p\\
&=L(f_{00}(X)g(X), Y=0\&T=1)\\
&+L(f_{11}(X)(1-g(X)), Y=1\&T=0)\\
&+L((f_{01}(X)+f_{11}(X))g(X), Y=1\&T=1)\\
&+L((f_{00}(X)+f_{11}(x))(1-g(X)), Y=0\&T=0)\\
&+L(g(X), T=1)
\end{aligned}
\label{LP}
\end{equation}

Given the results of stratification, the system only gives discounts on charging prices to the \textit{Incentive Charge} ECT-Hubs and avoids the \textit{Always Charge} ECT-Hubs, which is not considered in the previous causal inference-based method. In this way, the operators of ECT-Hubs can reduce costs and maximize profit. The experiments with real-world data in the next section prove the improvement of the proposed \textbf{ECT-Price} compared to the baselines.

\subsection{DRL-based BP Scheduling}

\begin{figure}[tb!]
\centerline{\includegraphics[scale=0.34]{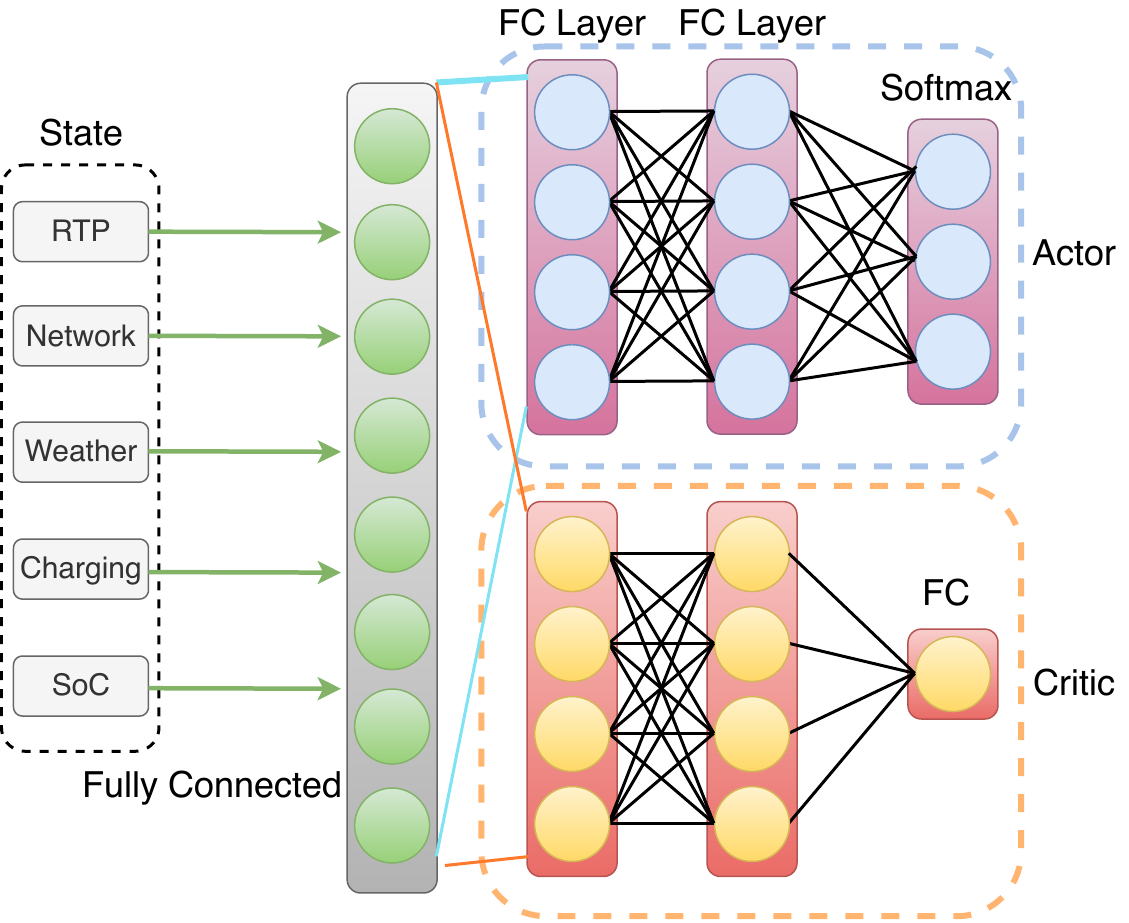}}
\caption{The architecture of ECT-DRL.}
\label{a2c}
\end{figure}

The proposed ECT-Hubs need to dynamically schedule the operation of BPs given the input of real-time price, weather conditions, network traffic, and charging price. As deep reinforcement learning has been shown to achieve great success in solving similar sequential decision problems with dynamic environments in other research areas, we propose a DRL-based method \textbf{ECT-DRL} to schedule the BP in the ECT-Hub. To this end, we first convert the scheduling problem into an RL task and then use a policy gradient algorithm named Actor-Critic method with \textit{Proximal Policy Optimization (PPO)} \cite{schulman2017proximal} algorithm to solve it. Since the state space in our problem is high-dimensional, we use Neural
Networks (NNs) to represent both the actor and critic components.

At first, we define the state of the environment at each time step $t$. According to the modeling of the system in Section III, the state $s_t$ can be represented as 

\begin{equation}
s_t=(\overrightarrow{RTP},\overrightarrow{weather}, \overrightarrow{traffic},\overrightarrow{SRTP},SoC)
\label{norm1}
\end{equation}
where $\overrightarrow{RTP}$ and $\overrightarrow{SRTP}$ are vectors to represent the real-time price from the power grid and charging price in a time period. $\overrightarrow{weather}$ and $\overrightarrow{traffic}$ are vectors to represent the weather conditions and network traffic in a time period. At last, the scalar $SoC$ is the current State of Charge of BP. Given the state and action of each time step, the environment gives back the reward $r_t$, which is calculated as Eq. \ref{obj}.

The network architecture of the model is shown in Fig.~\ref{a2c}. All the inputs are concatenated and fed into a fully connected layer, which is then fed to both the actor and critic network. For each input state $s_t$, the \textit{actor} component needs to output an action $a_t$, which determines the schedule of BP. There are three states for the BP, charge, discharge or not use. So we use three numbers $(0,1,2)$ to represent the operation of the BP. Unfortunately, the standard policy gradient algorithm can have great turbulence while training, which means the network may fall into a dead end and can never recover. To solve this problem, we use the \textit{Proximal Policy Optimization (PPO)} \cite{schulman2017proximal} algorithm to clip the size of each training step. Instead of the traditional objective $L^{P}_t(\theta)=\hat{\mathop{\mathbb{E}}}_t[\log\pi_{\theta}(a_t|s_t)\hat{A_t}]$, \textit{PPO} uses the clipped surrogate objective as

\begin{equation}
L^{clip}_t(\theta)=\hat{\mathop{\mathbb{E}}}_t[min(r_t(\theta)\hat{A_t}, clip(r_t(\theta),1-\epsilon,1+\epsilon)\hat{A_t}]
\label{norm1}
\end{equation}
where
\begin{equation}
r_t(\theta) = \frac{\pi_{\theta}(a_t|s_t)}{\pi_{\theta_{old}}(a_t|s_t)}
\label{norm1}
\end{equation}

$r_t(\theta)$ is the probability ratio of the new and old policy. $\hat{A_t}$ is the estimate of the \textit{advantage} function and $\epsilon$ controls the clip bounds. At last, the total loss function is the summation of this clipped \textit{PPO} objective and two additional terms:

\begin{equation}
L^{total}_t(\theta)=\hat{\mathop{\mathbb{E}}}_t[L^{clip}_t(\theta)-cMSE(V(s))]
\label{norm1}
\end{equation}
where $c$ is the coefficient of the second term, which is the mean square error of the value function $V(s)$. And the value function $V(s)$ is used to update the baseline network. Let $\alpha$ denote the learning rate, and the policy parameter $\theta$ can be updated as the gradient ascent of the total loss:
\begin{equation}
\theta \longleftarrow \theta+\alpha\sum_{t} \nabla_{\theta}L^{total}_t(\theta)
\label{norm1}
\end{equation}

\section{Performance Evaluation}

\subsection{Datasets and Evaluation Setup}

\begin{table*}[t!]
\caption{Performance evaluation of ECT-Price.}
\centering
\begin{tabular}{P{1cm}|P{0.8cm} P{0.8cm} P{0.8cm} P{0.8cm} | P{0.8cm} P{0.8cm} P{0.8cm} P{0.8cm} | P{0.8cm} P{0.8cm} P{0.8cm} P{0.8cm} }
 \hline
  & \multicolumn{4}{c|}{10\% Discount} & \multicolumn{4}{c|}{20\% Discount} & \multicolumn{4}{c}{30\% Discount} \\
 \hline
 \textbf{Methods} & None & Incentive & Always & Reward & None & Incentive & Always & Reward & None & Incentive & Always & Reward \\
 \hline
 OR & 2078 & 5936 & 412 & 5687 & 2078 & 5936 & 412 & 5439 & 2077 & 5937 & 412 & 5191 \\
 \hline
 IPS & 2079 & 5972 & 375 & 5727 & 2044 & 6071 & 311 & 5601 & 2058 & 6043 & 325 & 5329 \\
 \hline
 DR & 2053 & 6066 & 307 & 5830 & 2092 & 5801 & 533 & 5276 & 2083 & 5801 & 542 & 5014 \\
 \hline
 Ours & 1946 & 6398 & \textbf{82} & \textbf{6195} & 1990 & 6373 & \textbf{63} & \textbf{5963} & 1995 & 6355 & \textbf{76} & \textbf{5734} \\
 \hline
 \hline
  & \multicolumn{4}{c|}{40\% Discount} & \multicolumn{4}{c|}{50\% Discount} & \multicolumn{4}{c}{60\% Discount} \\
 \hline
 \textbf{Methods} & None & Incentive & Always & Reward & None & Incentive & Always & Reward & None & Incentive & Always & Reward \\
 \hline
 OR & 1938 & 5856 & 268 & 4975 & 1677 & 5789 & \textbf{23} & 4940 & 1510 & 5342 & \textbf{0} & 4437 \\
 \hline
 IPS & 2080 & 5977 & 369 & 4999 & 1829 & 5773 & 218 & 4751 & 2066 & 6067 & 293 & 4653 \\
 \hline
 DR & 2028 & 6117 & 281 & 5195 & 2061 & 6058 & 307 & 4876 & 1737 & 5755 & 88 & 4661 \\
 \hline
 Ours & 1973 & 6308 & \textbf{146} & \textbf{5462} & 1946 & 6397 & 83 & \textbf{5384} & 1969 & 6330 & 127 & \textbf{5072} \\
 \hline
\end{tabular}
\end{table*}

We implement a prototype and conduct extensive experiments on real-world datasets to evaluate the performance of the proposed methods. We acquire the weather information (wind speed and solar radiation) from NSRDB: National Solar Radiation Database \cite{weather}, and the RTP information from ENGIE Resources\cite{RTP}. The network traffic traces are obtained from a public dataset \cite{Modeling15-Chen}. We also collect three years of data from twelve charging stations in a campus, which contain more than $70,000$ rows of charging history. We label all the time slots with charging history as $Y=1$, and all the others are regarded as \textit{No Charge}. To further label \textit{Always Charge} and \textit{Incentive Charge} for the items with $Y=1$, we pre-trained a Neural Collaborative Filtering (NCF) \cite{NCF} model to generate the predicted ratings. Since \textit{Always Charge} tends to charge with more willingness, we label half of the items with the highest predicted ratings as \textit{Always Charge} and the remaining half as \textit{Incentive Charge}. 

We use three traditional uplift modeling methods: outcome regression (OR), inverse propensity scoring (IPS), and doubly robust (DR) estimators as baselines to evaluate our methods. All the baselines and the two tasks in ECT-Price use NCF as base models. All models are trained using Adam as the optimizer with a 0.01 learning rate and 1e-4 weight decay. ECT-DRL are trained using Adam as the optimizer with 1e-3 learning rate and 1e-4 weight decay. All of the models are trained on NVIDIA GeForce RTX 2080 Ti with batch size 64 and implemented in Python using PyTorch.

\begin{figure}[t]
  \centering
  \subcaptionbox{Station 1\label{fig3:a}}
  {\includegraphics[width=1.7in]{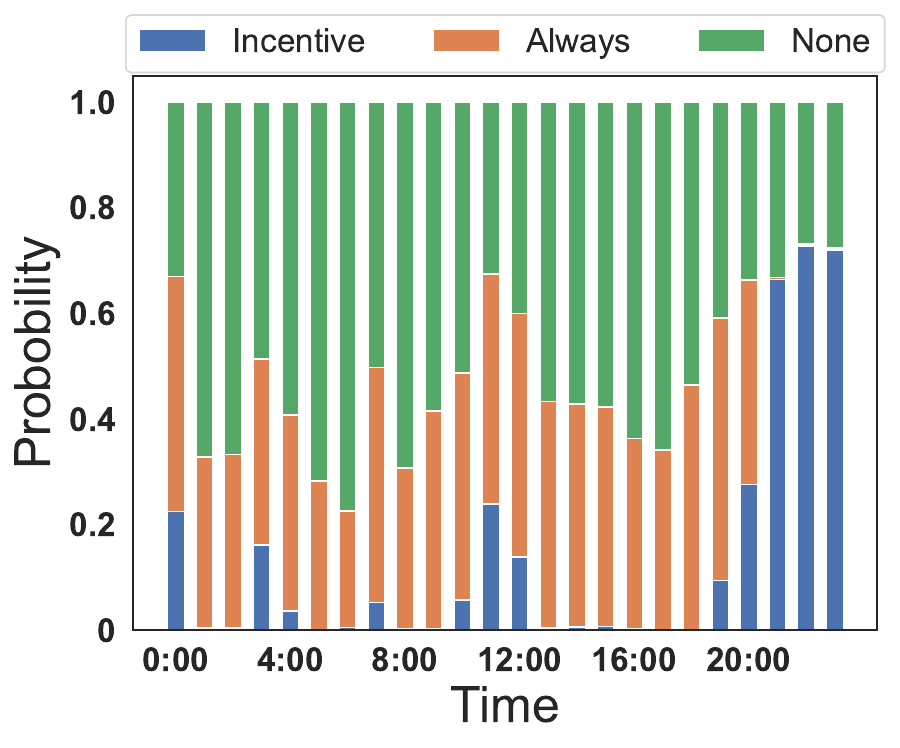}}\hspace{-0.2em}%
  \subcaptionbox{Station 2\label{fig3:a}}
  {\includegraphics[width=1.7in]{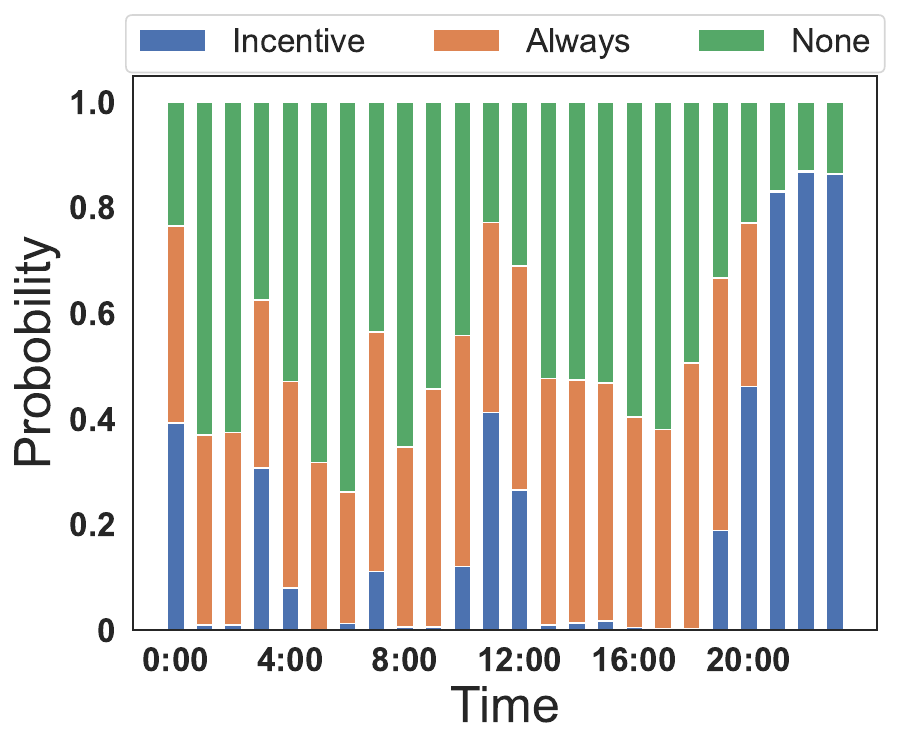}} 
  \subcaptionbox{Station 3\label{fig3:a}}
  {\includegraphics[width=1.7in]{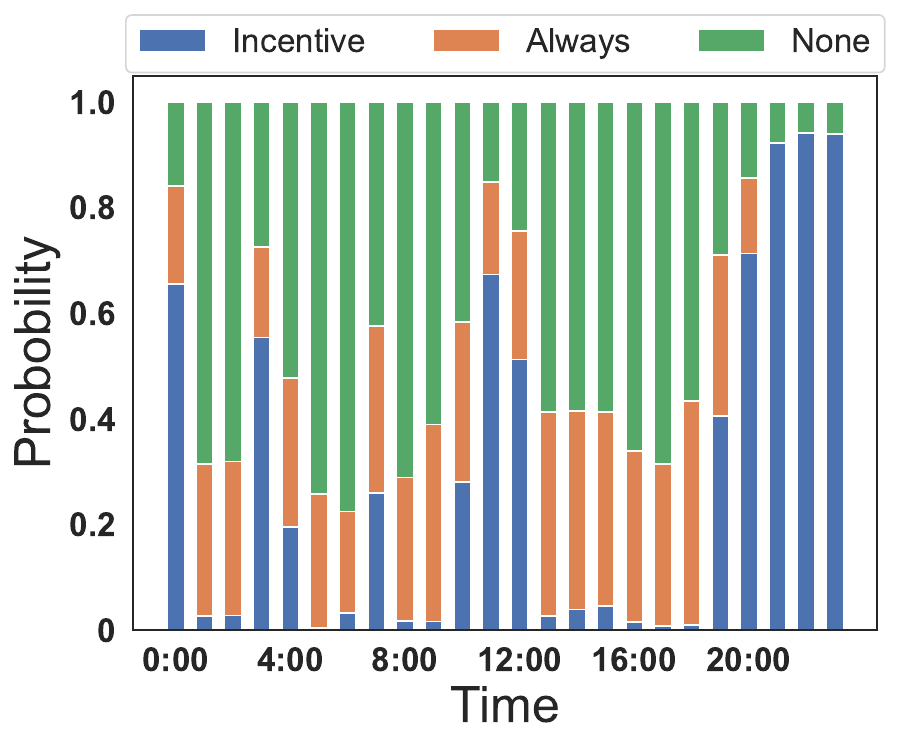}}\hspace{-0.2em}%
  \subcaptionbox{Station 4\label{fig3:a}}
  {\includegraphics[width=1.7in]{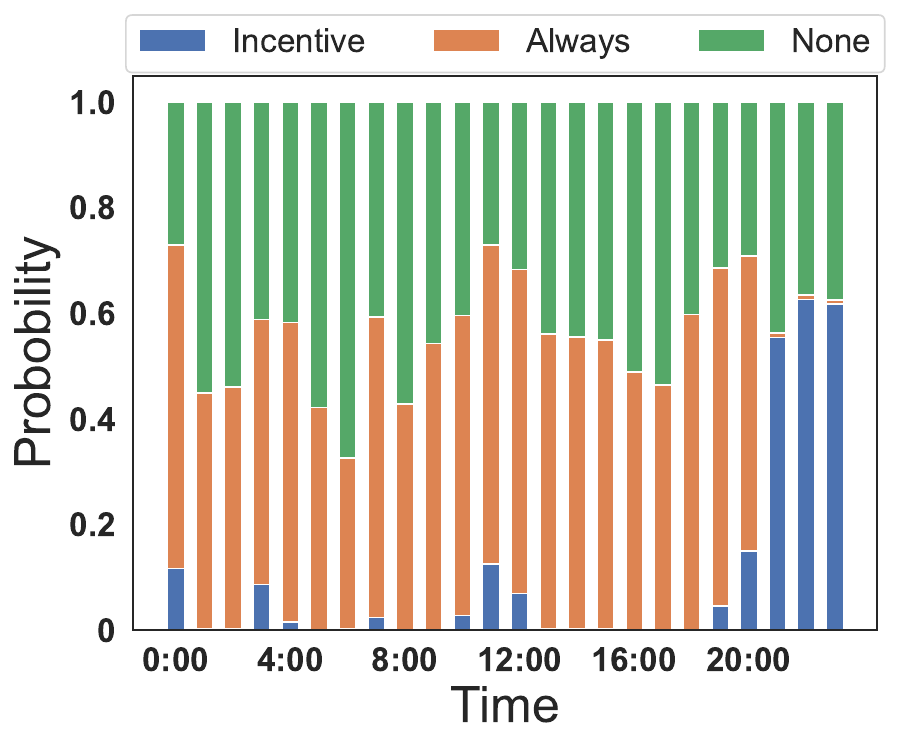}} 
  \caption{Strata prediction of four example stations.}
  \label{dis_str}
\end{figure}

\subsection{Performance Evaluation of ECT-Price}

At first, we compare the performance of ECT-Price with three baselines. We define the reward of \textit{Incentive Charge} as $-c$ when giving $c$ discount, the reward of \textit{Always Charge} as $1$, and the reward of \textit{No Charge} as $0$. We train the four methods with different percentages of discount using the collected datasets. Table II shows the number of different types of items that have been given discounts. When the discounts increase from $10\%$ to $60\%$, our method can consistently achieve the highest reward compared to the baselines, and the stratification results are the best most of the time.

\begin{figure}[t]
  \centering
  \subcaptionbox{00:00 to 06:00\label{fig3:a}}
  {\includegraphics[width=1.5in]{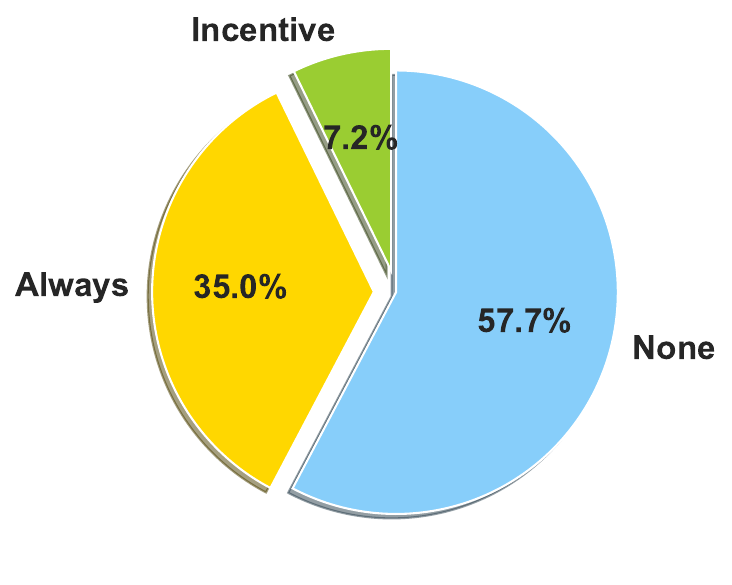}}\hspace{-0.2em}%
  \subcaptionbox{06:00 to 12:00\label{fig3:a}}
  {\includegraphics[width=1.5in]{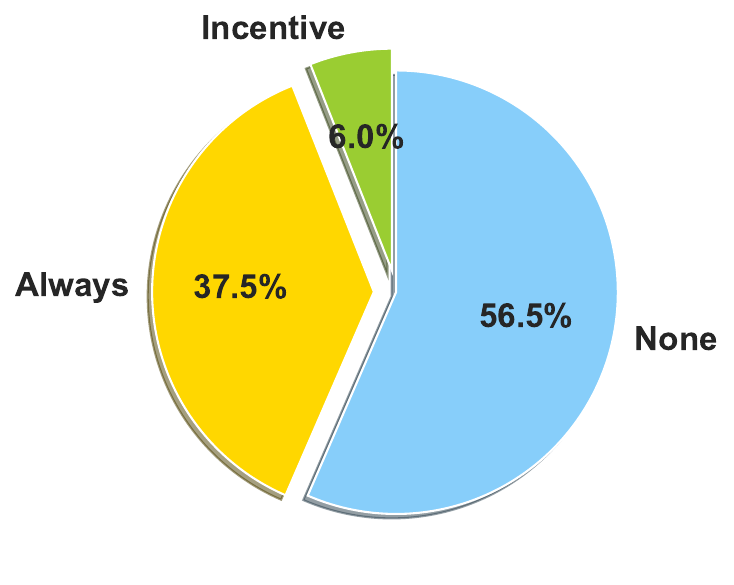}} 
  \subcaptionbox{12:00 to 18:00\label{fig3:a}}
  {\includegraphics[width=1.5in]{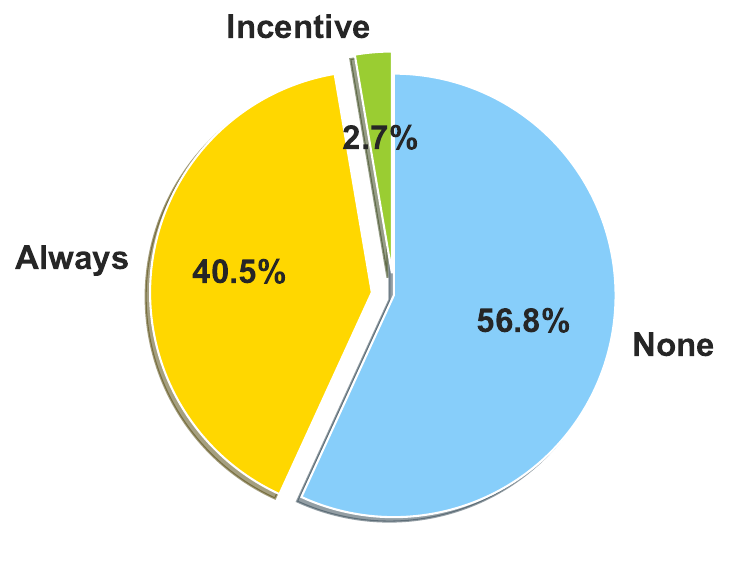}}\hspace{-0.2em}%
  \subcaptionbox{18:00 to 24:00\label{fig3:a}}
  {\includegraphics[width=1.5in]{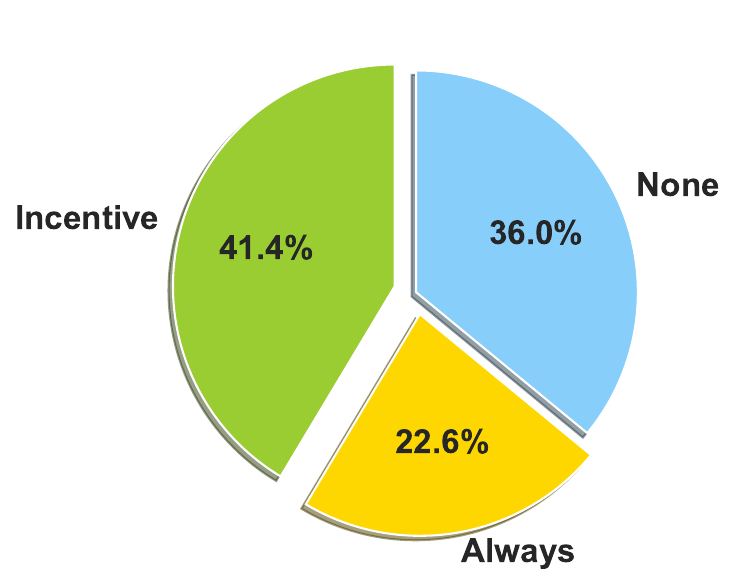}} 
  \caption{Strata distribution of four periods. \textit{Incentive Charge} tends to appear more at the last time period.}
  \label{pie}
\end{figure}

To further illustrate the distribution of different strata, we show the stratification results of four example charging stations in Fig.~\ref{dis_str}. For all the examples, \textit{Incentive Charge} tends to appear at night, and \textit{Always Charge} appears more at other time slots. To further prove this observation, we divide the time slots into four periods and plot pie charts to see the distribution, as shown in Fig.~\ref{pie}. More items become \textit{Incentive Charge} during the period of $18:00-24:00$,  which means that the ECT-Hub should give more discounts during this period to attract more EVs charging and make more profit.

\subsection{Performance Evaluation of ECT-DRL}

\begin{figure}[t]
  \centering
  \subcaptionbox{Hub 1\label{fig3:a}}
  {\includegraphics[width=1.7in]{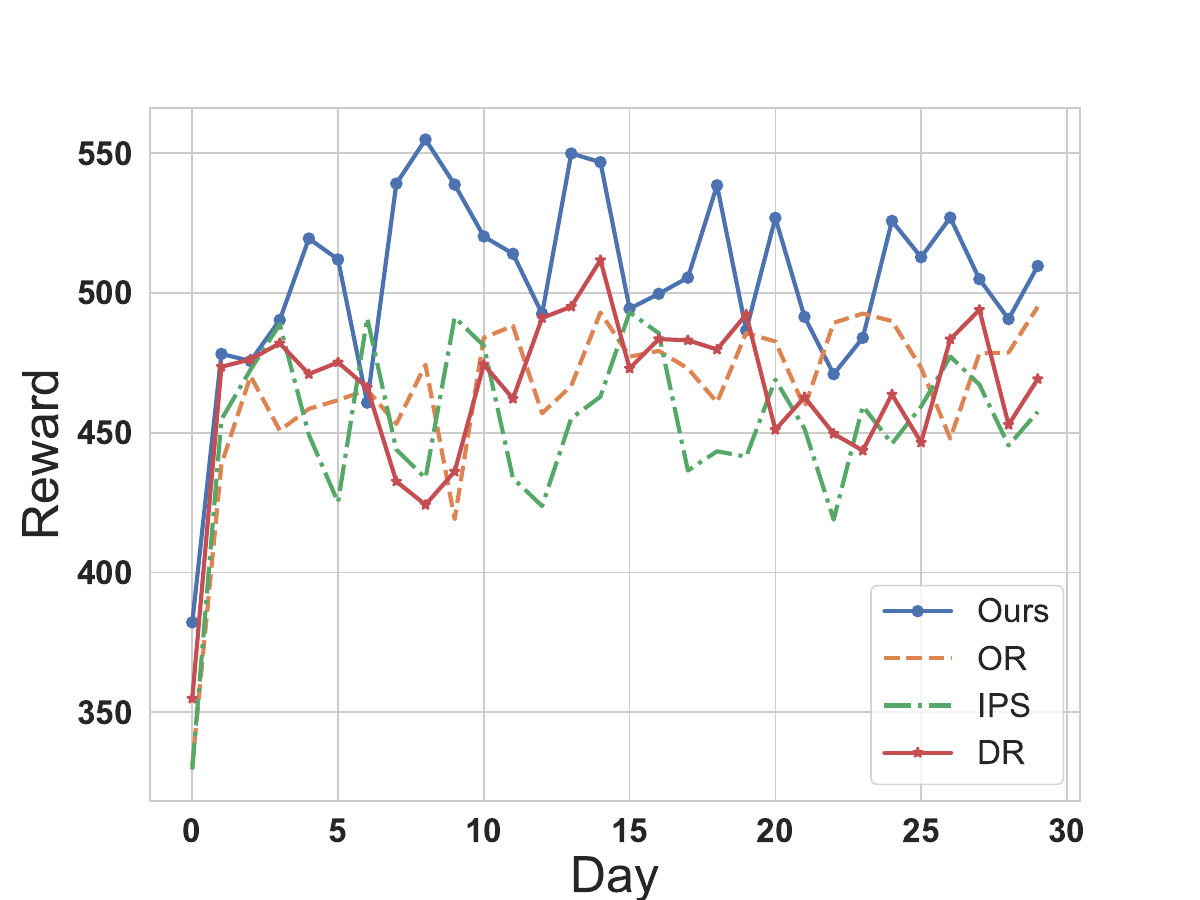}}\hspace{-1.2em}%
  \subcaptionbox{Hub 2\label{fig3:a}}
  {\includegraphics[width=1.7in]{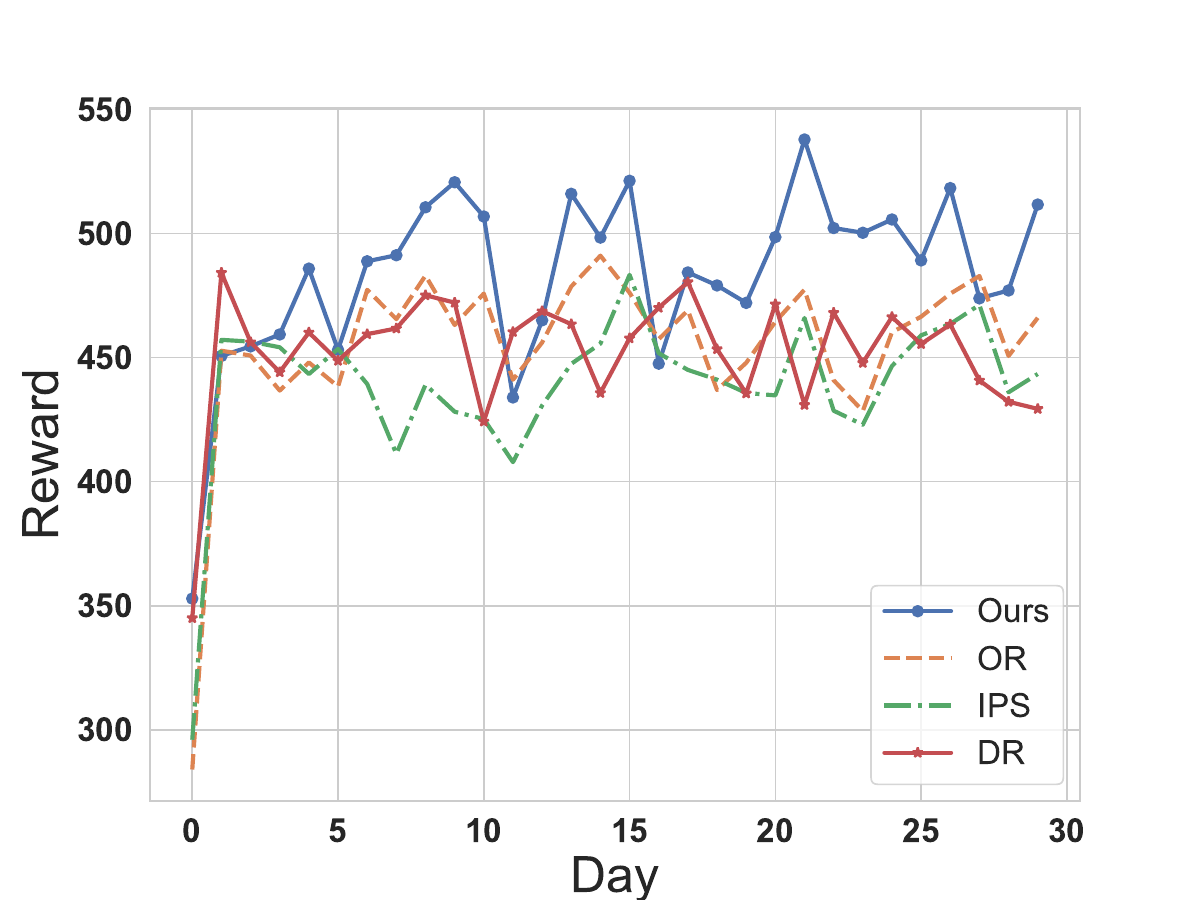}} 
  \subcaptionbox{Hub 3\label{fig3:a}}
  {\includegraphics[width=1.7in]{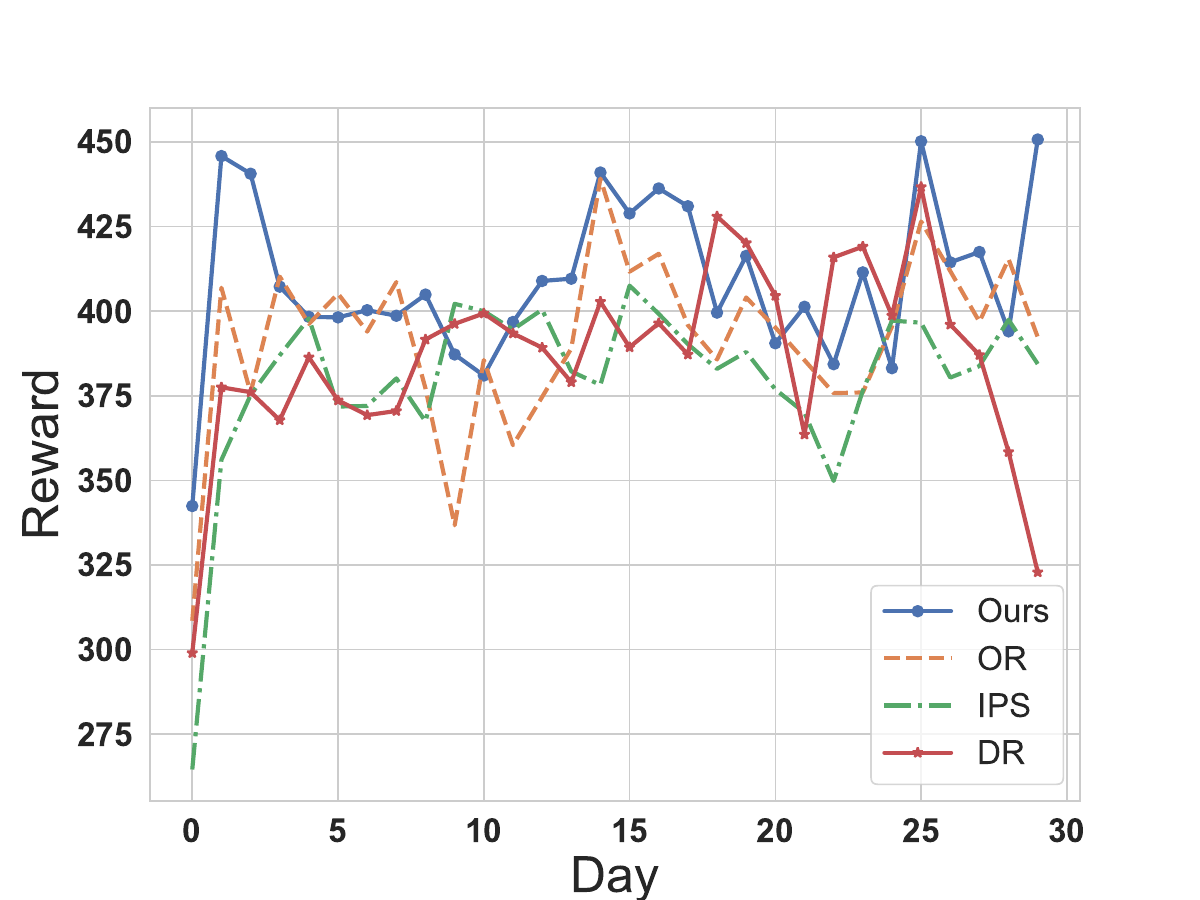}}\hspace{-1.2em}%
  \subcaptionbox{Hub 4\label{fig3:a}}
  {\includegraphics[width=1.7in]{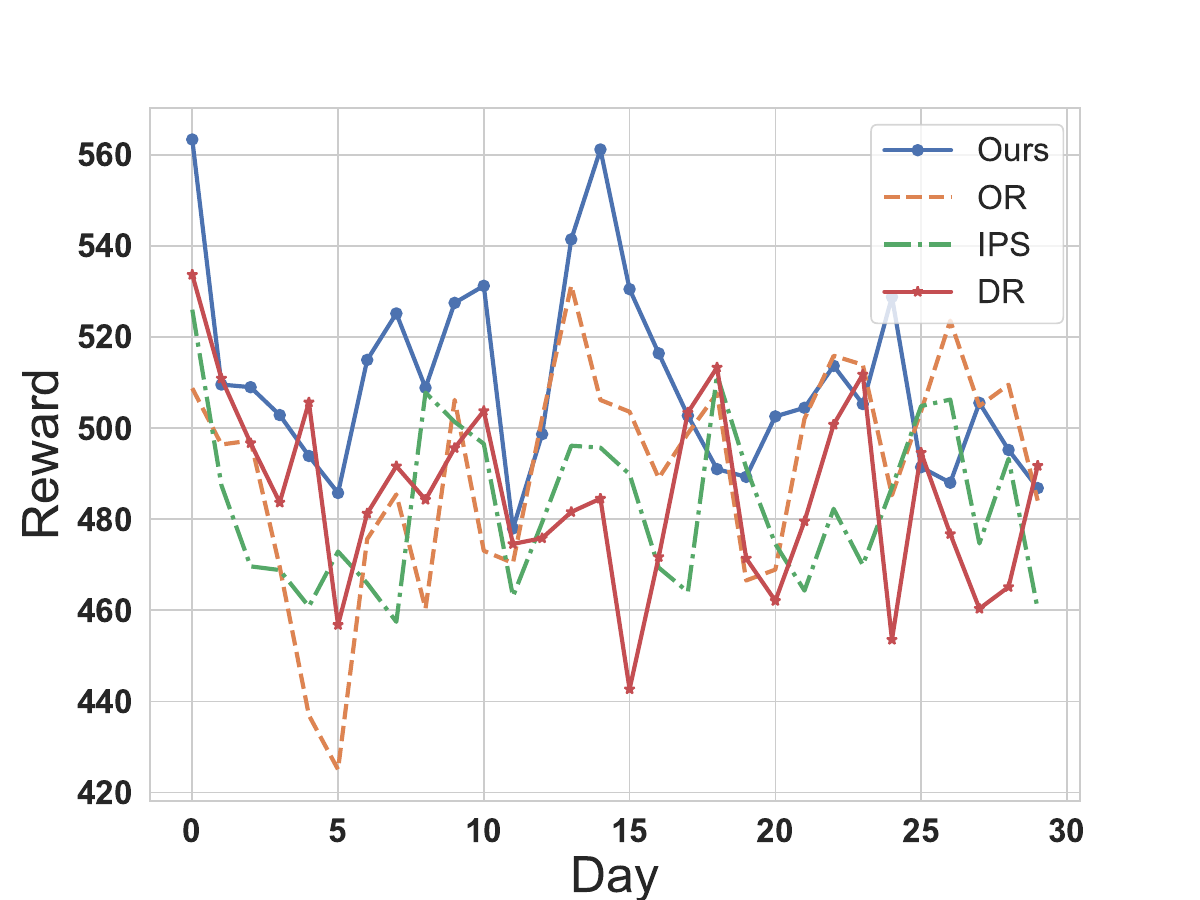}} 
  \caption{Total reward of four example hubs. Our methods achieve the best average reward compared to baselines.}
  \label{DRL}
\end{figure}

At last, we evaluate the performance of ECT-DRL. We denote 30 days as one episode in our environment. We train our system for 500 episodes and test it for 100 episodes. The operation cost of BP for each time slot is set to 0.01, and the SoC of the BP is set to a random percent at the start of each episode. For each ECT-Hub, we train four models with different inputs of the charging price, which are based on the prediction results of ECT-Price and three baselines. All the other inputs, such as weather information and network traffic traces, remain the same for the four models. Fig.~\ref{DRL} shows the average daily reward. Due to space limitations, we only plot the results of four example ECT-Hubs. The results show that our method has the best performance most of the time and achieves a higher average reward than the baselines. Table III shows the results of all the twelve ECT-Hubs in the datasets. In summary, our method achieves the highest average daily rewards for all the ECT-Hubs, which shows the effectiveness of our methods in maximizing the profit of ECT-Hubs.

\begin{table*}[t!]

\centering
\caption{Average daily rewards for 12 ECT-Hubs.}
\label{tab:average-reward}
\begin{tabular}{P{1cm}|cccccccccccc}
\hline
Methods & Hub1 & Hub2 & Hub3 & Hub4 & Hub5 & Hub6 & Hub7 & Hub8 & Hub9 & Hub10 & Hub11 & Hub12 \\
\midrule 
\hline
OR & 529.57 & 453.08 & 385.44 & 498.88 & 535.48 & 483.43 & 488.83 & 514.69 & 332.33 & 519.09 & 473.27 & 534.02 \\
\hline
IPS & 498.63 & 440.21 & 373.04 & 486.07 & 526.7 & 459.37 & 478.72 & 498.03 & 305.15 & 514.06 & 462.06 & 534.27 \\
\hline
DR & 535.58 & 449.32 & 384.31 & 497.78 & 535.05 & 474.18 & 492.32 & 515.61 & 325.05 & 511.27 & 459.86 & 542.06 \\
\hline
Ours & \textbf{565.19} & \textbf{488.05} & \textbf{400.41} & \textbf{510.22} & \textbf{566.03} & \textbf{496.36} & \textbf{512.98} & \textbf{533.42} & \textbf{352.29} & \textbf{540.86} & \textbf{499.76} & \textbf{563.12} \\
\hline
\end{tabular}
\end{table*}
\section{Related Work}

\subsection{Base Stations Energy Management}

Current research mainly focuses on the energy consumption patterns of 5G base stations and the optimization of the behavior of base station energy storage batteries. Tang \textit{et al.} \cite{tang2020shiftguard} proposed a backup power deployment framework to reduce the deployment cost while ensuring the normal operation of 5G base stations. The authors consider the power demand differences in spatial and temporal relationships between base stations as well as network reliability and practical deployment constraints. They further proposed a deep reinforcement learning-based approach to address challenges in optimizing BESS discharge/charge scheduling within a heterogeneous network (HetNet) architecture \cite{tang2021reusing}. Their method considers practical BS power supply/demand and battery specifications and repurposes the backup batteries as the distributed BESS. The result demonstrates that using distributed BESS and their DRL-based scheduling approach can significantly reduce the mobile operator's demand charge and result in substantial yearly operational expenditure savings for a large number of 5G BSs. Most of the current works only consider the base station as the consumer of energy from the storage battery. However, the capacity of the storage batteries and renewable power generation could be well beyond the demand of the base station and not be fully utilized. Our proposed ETC-Hubs additionally take into account the function of charging for EVs, which makes more profit for the operators while maintaining the basic functions of base stations.

\subsection{Electric Vehicles Charging Behavior}
Electric vehicles have become an important part of the modern power grid. Many works have explored the charging behavior of electric vehicles and investigated their effect on the power system. The model proposed in \cite{bae2011spatial} captures the spatial and temporal dynamics of EV charging demand at a highway rapid charging station, using fluid dynamic traffic modeling and M/M/s queuing theory. Hung \textit{et al.} \cite{hung2022novel} proposed a data-driven method for the location-routing problem of charging infrastructure under fixed charging station numbers and random charging demands. Some works \cite{zhao2021joint}\cite{wang2022towards} also focus on the urban vehicle-sharing system and propose methods to balance the user demand and charging requirements.

There have been some studies on the electric vehicle battery swapping and deployment of Battery Swapping Stations (BSS). However, BSS is not widely applied due to efficiency, cost, and battery specifications. Some initiatives, such as Tesla's abandonment of its pilot program for switching stations in favor of a major push into plug-in charging, demonstrate the limitations of this technology. Compared to electric vehicles, battery swap stations are more prevalent in the electric two-wheeler domain and are typically deployed near base stations. Considering the challenges of EV mileage, cost, and BSS scale, widespread use in everyday electric vehicle usage is currently not deemed practical \cite{wu2017optimization,kang2015centralized,ding2021integrated,ahmad2020battery}. Wang \textit{et al.} \cite{wang2016electric} proposed an optimization model for charging infrastructure deployment based on heuristic algorithms. Existing EV charging models are vehicle-based and designed for optimizing the deployment location of charging piles, while our proposed ECT-Hubs are transformed from the current 5G base stations. In our design, we consider the charging needs of different vehicles in the case of fixed charging station locations, distinguish between just-demand users and incentive users, and use this to set different electricity prices to maximize profit.

\subsection{Causal Inference}
Causal inference enables researchers to draw conclusions about causal relationships between variables using assumptions, study designs, and estimation strategies based on data. This process aids in understanding underlying mechanisms in complex systems and making informed decisions \cite{wu2022instrumental}.
Kuang \textit{et al.} \cite{kuang2017estimating} propose a data-driven Differentiated Confounder Balancing (DCB) algorithm that effectively reduces confounding bias in high-dimensional settings. Extensive experiments on synthetic and real datasets demonstrate the superior performance of the DCB algorithm. Xu \textit{et al.} \cite{xu2020learning} propose a deep feature instrumental variable regression (DFIV) method to address the case where relations between instruments, treatments, and outcomes may be nonlinear. Although causal inference has been widely used in e-commerce and has shown great success, there is still no previous work that utilizes causal inference to determine the electricity price. In this work, we use price discounts as a treatment to differentiate between users who simply require charging and those who are incentivized. The experiment results show that our method can achieve the best profit compared to the baselines.

\section{Conclusion}

In this work, we introduced an innovative framework called the Energy-Communication-Transportation Hub (ECT-Hub), which integrates energy storage, communication services, and transportation capabilities. ECT-Hub provides a base station-centric design that enables synergies between energy storage and EV charging, utilizing the excess energy storage resources in the hub to guarantee the reliable operation of the communication system and provide extra charging services for EVs. Benefiting from this, the operators can make more profit while the customers can also have more convenient charging services. Our framework incorporates weather information, communication traffic, and EVs charging history to optimize the scheduling of energy storage batteries at ECT-Hubs. Deep reinforcement learning-based methods are used to solve this optimization problem. Additionally, we have designed an incentive mechanism using causal inference algorithms to differentiate EV charging demands and set the corresponding pricing. The profit is maximized by avoiding unnecessary electricity price discounts. We implemented a prototype and the experiments with real-world data demonstrated the effectiveness of the methods and showed the feasibility of the base-station-centric ECT-Hub design.


\newpage

\balance

\bibliographystyle{IEEEtran}
\bibliography{main.bib}

\end{document}